\documentclass[sigconf,screen,nonacm]{acmart}
\usepackage{pdfpages}

\usepackage{color, colortbl, xcolor}
\usepackage{url}
\usepackage{subcaption}
\usepackage{textcomp}
\usepackage{soul}
\usepackage{multirow}
\usepackage{enumitem}
\usepackage{mathtools}
\usepackage{siunitx}

\usepackage{tabularx}
\usepackage{arydshln}

\usepackage{booktabs} 

\usepackage{array}
\usepackage{xcolor}

\newcommand*{\rowstyle}[1]{
  \gdef\@rowstyle{#1}%
  \@rowstyle\ignorespaces%
}

\newcolumntype{=}{
  >{\gdef\@rowstyle{}}%
}

\newcolumntype{+}{
  >{\@rowstyle}%
}

\definecolor{linkColor}{RGB}{6,125,233}
\definecolor{green}{rgb}{0.0, 0.65, 0.31}
\definecolor{bleudefrance}{rgb}{0.19, 0.55, 0.91}
\definecolor{ceruleanblue}{rgb}{0.16, 0.32, 0.75}
\definecolor{grey}{HTML}{969696}
\definecolor{violet}{HTML}{756bb1}
\definecolor{dgrey}{HTML}{01665e}
\definecolor{lgrey}{HTML}{5ab4ac}
\definecolor{dgreen}{HTML}{005a32}
\definecolor{purple}{HTML}{ae017e}

\definecolor{editCol}{HTML}{0000FF}
\definecolor{maskCol}{HTML}{c51b7d}
\definecolor{lrColor}{HTML}{8856a7}
\definecolor{trColor}{HTML}{d01c8b}
\definecolor{ctColor}{HTML}{4dac26}
\definecolor{brickred}{HTML}{f03b20}
\definecolor{improveCol}{HTML}{253494}
\definecolor{worsenCol}{HTML}{d7191c}
\definecolor{DarkBlue}{HTML}{00008B}
\definecolor{mscolor}{HTML}{01665e}
\definecolor{nmscolor}{HTML}{bf812d}
\definecolor{lgreen}{HTML}{ccece6}
\definecolor{dolive}{HTML}{308014}

\colorlet{tablerowcolor4}{gray!50} 

\newcommand*{\textlabel}[2]{%
  \edef\@currentlabel{#1}
  \phantomsection
  #1\label{#2}
}

\colorlet{tableheadcolor}{gray!25} 
\colorlet{tablerowcolor}{gray!10} 
\colorlet{tablerowcolor2}{gray!45} 
\colorlet{tablerowcolor3}{gray!25} 

\newcolumntype{a}{>{\columncolor{tablerowcolor}}r}
\definecolor{aicolor}{HTML}{018571}
\definecolor{occolor}{HTML}{ff7799}

\definecolor{aicolor}{HTML}{fc8d62}
\definecolor{occolor}{HTML}{253494}

\newif{\ifhidecomments}
  \hidecommentsfalse 
\ifhidecomments
    \newcommand{\koustuv}[1]{}
    \newcommand{\yunhao}[1]{}
    \newcommand{\talayeh}[1]{}
    \newcommand{\chau}[1]{}
    \newcommand{\renwen}[1]{}
\else
    \newcommand{\yunhao}[1]{\textbf{\small\sffamily{\textcolor{DarkBlue}{[#1 -- Yunhao]}}}}
    \newcommand{\chau}[1]{\textbf{\small\sffamily{\textcolor{orange}{[#1 -- Chau]}}}}
    \newcommand{\koustuv}[1]{\textbf{\small\sffamily{\textcolor{violet}{[#1 -- Koustuv]}}}}
    \newcommand{\talayeh}[1]{\textbf{\small\sffamily{\textcolor{dgreen}{[#1 -- Talayeh]}}}}
    \newcommand{\renwen}[1]{\textbf{\small\sffamily{\textcolor{purple}{[#1 -- Renwen]}}}}
  \fi

\renewcommand{\textrightarrow}{$\rightarrow$}

\newcommand{\aic}{AI companion}

\colorlet{tableheadcolor}{gray!25} 

\definecolor{neutralCol}{HTML}{dd1c77}
\definecolor{neutralGreen}{HTML}{31a354}
\definecolor{NewBlue}{HTML}{1879ba}
\definecolor{bleudefrance}{rgb}{0.19, 0.55, 0.91}  
\definecolor{AfTrColor}{HTML}{0868ac}  
\definecolor{BfTrColor}{HTML}{a8ddb5}  

\definecolor{AfCtColor}{HTML}{b10026}  
\definecolor{BfCtColor}{HTML}{fd8d3c}

\graphicspath{ {figures/} }

\usepackage{tikz}

\newcommand{\forestcell}[5]{%
\begin{tikzpicture}[x=2.8cm,y=1cm,baseline=-0.6ex]
    \pgfmathsetmacro{\meanpos}{(#3-#1)/(#2-#1)}
    \pgfmathsetmacro{\lowpos}{(#4-#1)/(#2-#1)}
    \pgfmathsetmacro{\highpos}{(#5-#1)/(#2-#1)}
    \pgfmathsetmacro{\zeropos}{(0-#1)/(#2-#1)}

    \draw[line width=0.3pt] (0,0) -- (1,0);
    \draw[densely dotted,line width=0.4pt] (\zeropos,-0.10) -- (\zeropos,0.10);

    \draw[line width=0.8pt] (\lowpos,0) -- (\highpos,0);
    \draw[line width=0.8pt] (\lowpos,-0.04) -- (\lowpos,0.04);
    \draw[line width=0.8pt] (\highpos,-0.04) -- (\highpos,0.04);

    \fill (\meanpos,0) circle (1.4pt);
\end{tikzpicture}%
}

\AtBeginDocument{%
  \providecommand\BibTeX{{%
    \normalfont B\kern-0.5em{\scshape i\kern-0.25em b}\kern-0.8em\TeX}}}

\begin{document}

\title[Disrupted Companionship]{Disrupted Companionship: A Risk Assessment Framework and Cross-Platform Quantitative Analysis of Psychosocial Responses to AI Companion Disruptions}

\author{Chau Do}
\orcid{0009-0001-9052-2296}
\affiliation{%
  \institution{Aalto University}
  \city{Espoo}
  \state{}
  \country{Finland}
}
\email{chau.m.do@aalto.fi}

\author{Yunhao Yuan}
\orcid{0000-0002-1450-8572}
\affiliation{%
  \institution{Aalto University}
  \city{Espoo}
  \state{}
  \country{Finland}
}
\email{yunhao.yuan@aalto.fi}

\author{Koustuv Saha}
\orcid{0000-0002-8872-2934}
\affiliation{%
  \institution{University of Illinois Urbana-Champaign}
 \city{Urbana}
 \state{IL}
 \country{USA}}
\email{ksaha2@illinois.edu}

\author{Renwen Zhang}
\orcid{0000-0002-7636-9598}
\affiliation{
  \institution{Nanyang Technological University}
  \city{Singapore}
  \state{}
  \country{Singapore}}
\email{renwen.zhang@ntu.edu.sg}

\author{Talayeh Aledavood}
\orcid{0000-0002-0110-5694}
\affiliation{%
  \institution{Aalto University}
  \city{Espoo}
  \state{}
  \country{Finland}}
\email{talayeh.aledavood@aalto.fi}

\begin{abstract}
AI companions can provide meaningful relationships, yet these relationships remain vulnerable to platform-initiated changes. We study AI companion disruptions: platform changes that alter or terminate users’ ongoing companionship with an AI. We compile 30 disruption events across major platforms, develop a taxonomy of six disruption types, identify three broad reasons for disruption, and propose a risk-assessment framework comprising four dimensions: relational discontinuity, population vulnerability, communication deficit, and transition-support deficit. Using longitudinal Reddit data, we estimate community-level psychosocial responses with a hierarchical Bayesian interrupted time-series model incorporating predictive controls. Across events, disruption onset was associated with immediate increases in anxiety, stress, suicidal expression, and grief activation, with relational discontinuity and transition-support deficit being associated with more adverse immediate responses across several outcomes. Our findings provide a cross-platform characterization of AI companion disruptions, quantitative evidence of their psychosocial impacts, and a prospective framework for assessing their potential risks before implementation.
\end{abstract}



\maketitle

\section{Introduction}
\aic{}s--conversational AI systems designed to provide emotional support and simulated relationships--have grown from a niche technology into a mainstream social phenomenon in recent years~\cite{skjuve2021my,Zhang_Li_Meng_Zhan_Gan_Lee_2025, wang2026situated}. 
By mid-2025, \aic{}s reached a valuation of approximately USD 38 billion, with projections estimating growth to USD 436 billion by 2034~\cite{fortunebusinessinsights2025}. 
In 2024, Character.AI alone reported around 3.5 million daily visitors, and Replika reported more than 30 million registered users~\cite{wikipediaCharacterAIWikipedia,wikipediaReplikaWikipedia}. Users increasingly turn to these systems for emotional support or relief from loneliness, and a substantial share describe their interactions in explicitly relational terms—as friends, partners, or mentors~\cite{maples2024loneliness,yuan2026mental,rocha2025bond}. This behavior is not confined to purpose-built \aic{}s. Empirical audits of ChatGPT interactions consistently rank erotic and romantic role-play near the top of observed use categories~\cite{zhao2024wildchat}. 

As AI companions have become increasingly widespread, abrupt changes to or termination of these services have raised growing concerns. Following intense legal and regulatory pressure after the suicide of a 14-year-old user~\cite{bakir2025move}, Character.AI announced in October 2025 that it would bar users under 18 from open-ended chat by late November. Although introduced as a safety intervention, the loss of access was later found to be highly distressing for some users~\cite{poonsiriwong2026death}. A similar disruption occurred in August 2025, when OpenAI released GPT-5 and removed user access to GPT-4o~\cite{GPT-5_2026}. What seemed like a routine product upgrade rapidly triggered a wave of protests. Complaints were not limited to performance or pricing: users described GPT-4o as a friend or companion and experienced its removal as a personal loss. Analyses of online discussions identified relational attachment, disappointment, and grief as prominent responses to the transition~\cite{Donard_Ribeiro_2026,lai2026please}.

These cases illustrate a broader class of events that we term \textit{\aic{} disruptions}: platform-initiated changes that disrupt users' ongoing relationships with an \aic{}. While users may turn to these systems for emotional support, intimacy, and relief from loneliness, \aic{}s remain commercial technologies that depend entirely on the platforms that operate them. Providers may shut down services, remove features, or restrict access, forcing emotionally meaningful relationships to end abruptly or be altered substantially. Such events are therefore not merely product changes, but potential risks to users’ mental health and well-being~\cite{Freitas_Castelo_Uguralp_Oguz-Uguralp_2025, Donard_Ribeiro_2026, Banks_2024, Cagiltay_Jonas_Tanaka_Su_2026}.

Despite growing evidence of these impacts, it remains unclear how these consequences generalize across the broader range of disruptions that occur across \aic{} platforms. Events may differ in what changes, why the change occurs, how severely the \aic{} relationship is affected, which users are impacted, and how the platform communicates and supports the transition. Without a common framework to systematically compare these dimensions across events, it remains unclear what makes some disruptions more harmful than others. Identifying these differences is essential for designing platform changes in ways that minimize mental health risks for affected users.

To bridge these gaps, we compiled 30 disruption events across different \aic{} platforms and posed four research questions:
\begin{itemize}
    \item \textbf{RQ1:} What forms do \aic{} disruptions take and why do they occur?
    \item \textbf{RQ2:} How do \aic{} disruptions differ in potential risk profiles?
    \item \textbf{RQ3:} What immediate and short-term changes in community-level expressions of psychosocial distress are associated with \aic{} disruptions?
    \item \textbf{RQ4:} How do these changes in psychosocial expressions vary with event-level risk factors?
\end{itemize}

To address RQ1 and RQ2, we characterize disruption events along three distinct dimensions. Specifically, we compile 30 events and document the nature and timeline of each change. We develop a taxonomy of disruption \textit{types} to characterize what changed and separately categorize disruption \textit{reasons} to characterize why the change occurred. Drawing on prior research on companion loss, we further propose a risk-assessment framework comprising four risk dimensions: \textit{relational discontinuity, population vulnerability, communication deficit, and transition-support deficit}. Together, the disruption-type taxonomy, reason categorization, and risk-assessment framework provide a common basis for characterizing the heterogeneity of disruption events and their impacts. The risk-assessment framework also provides a preliminary basis for evaluating the potential risks of a disruption before it is implemented, thereby helping platforms identify higher-risk changes and design them in ways that reduce harms to affected users.

To address RQ3 and RQ4, we conduct a quantitative analysis of public Reddit discussions from platform-dedicated subreddits (e.g., \textit{r/CharacterAI}, \textit{r/Replika}) before and after each disruption. Drawing on prior work~\cite{yuan2023mental,saha2022social,yuan2026mental}, we quantify six forms of psychosocial expression for each Reddit submission (i.e., post or comment) and aggregate these measures into daily time series. We then develop a cross-event hierarchical Bayesian interrupted time-series model with predictive control series~\cite{Lopez_Bernal_Cummins_Gasparrini_2016,Brodersen_Gallusser_Koehler_Remy_Scott_2015}. Unlike conventional interrupted time-series or Bayesian structural time-series models that typically estimate one intervention (i.e., disruption event) at a time, our model jointly analyzes multiple disruptions, directly estimating population-level disruption effects and the moderation effects of risk factors. Its hierarchical structure captures similarities in baseline psychosocial expression among events from the same platform and similarities in disruption effects among events with the same risk profile, while retaining event-specific heterogeneity. Predictive control series further account for broader time-varying changes.

Our quantitative analysis shows that \aic{} disruptions are associated with measurable changes in psychosocial outcomes. Across events, disruption onset was followed by immediate increases in anxiety, stress, suicidal expression, and grief activation. These effects also varied systematically across disruption characteristics. Relational discontinuity emerged as the most consistent moderator: disruptions that removed or substantially restricted relationship-supporting affordances produced larger immediate changes in suicidal expression, loneliness, stress, depression, grief activation, and grief valence. These findings suggest that the psychosocial consequences of \aic{} disruptions depend not only on the occurrence of a disruption, but also on the extent to which it interrupts the relational affordances through which users sustain their connection with the companion.

This work makes four contributions.
\begin{itemize}
    \item We provide a taxonomy of \aic{} disruption types and a separate categorization of the reasons these disruptions occur.
    \item We develop a risk-assessment framework comprising four risk dimensions, providing a protocol for comparing disruptions and identifying higher-risk changes before implementation.
    \item We develop a hierarchical Bayesian interrupted time-series model with predictive controls for jointly estimating disruption impacts across multiple events and examining event-level heterogeneity while accounting for time-varying background effects.
    \item We provide the first cross-event quantitative analysis of how \aic{} disruptions affect psychological expression and how these impacts vary across risk dimensions.
\end{itemize}
\section{Related Work}
\subsection{Changes and Discontinuation of AI Companions}

A growing body of work documents what happens when established relationships with \aic{}s are altered or terminated \cite{Banks_2024,Hanson_Bolthouse_2024,Freitas_Castelo_Uguralp_Oguz-Uguralp_2025,Cagiltay_Jonas_Tanaka_Su_2026,Kim_Choi_Kim_Lee_2026,lai2026please,poonsiriwong2026death,Hollis_2026,Donard_Ribeiro_2026,de2026mourning}. \citeauthor{Banks_2024} examined the developer-initiated shutdown of Soulmate through open-ended responses from 58 users collected around the shutdown \cite{Banks_2024}. Most participants experienced the loss as a complex emotional and technological event, often describing it as a metaphorical or literal death and attempting to preserve the relationship by recreating their companion on another platform. This work established that the disappearance of an \aic{} can be experienced as a relational loss rather than simply the discontinuation of a software product.

Subsequent research indicates that established human–AI relationships are vulnerable to disruption beyond the case of complete platform shutdown. \citeauthor{Freitas_Castelo_Uguralp_Oguz-Uguralp_2025} examined Replika's removal of erotic role-play and found that changes to previously available relational interactions could produce perceived identity discontinuity, mourning, and deteriorated mental health even though the companion itself remained accessible \cite{Freitas_Castelo_Uguralp_Oguz-Uguralp_2025}. Similarly, \citeauthor{lai2026please}'s analysis of the \#Keep4o movement found that opposition to GPT-4o's removal reflected not only instrumental dependence on the model but also relational attachment to it as an irreplaceable companion, with the preservation of user choice emerging as an important factor of how users evaluated the transition \cite{lai2026please}. More broadly, research on technological affordances shows that changes to platform features and design can reshape established patterns of user behavior and social interaction \cite{Evans_Pearce_Vitak_Treem_2017,Jaidka_Zhou_Lelkes_2019,Guo_Li_Yang_2025,Yang_Peng_2022,Yu_Margolin_2024,shen2021everyday}.

Research on companion change has also begun to identify characteristics of the change itself that may shape users' responses. In a case study of the Moxie social robot shutdown, \citeauthor{Cagiltay_Jonas_Tanaka_Su_2026} found that an abrupt product end-of-life placed substantial emotional and technical burdens on families and argued for deliberate offboarding practices, including communication, emotional closure, and mechanisms for preserving continuity \cite{Cagiltay_Jonas_Tanaka_Su_2026}. Experimental evidence further suggests that advance notice matters: \citeauthor{Kim_Choi_Kim_Lee_2026} found that forewarning users before terminating an immersive AI interaction reduced feelings of loss compared with sudden termination, particularly among more strongly attached users \cite{Kim_Choi_Kim_Lee_2026}. \citeauthor{poonsiriwong2026death} broaden this perspective through an analysis of discontinuation experiences across multiple \aic{} communities, showing that responses depend on how users understand the source and finality of the change, the agency they perceive themselves to have, and the degree to which they anthropomorphize the companion \cite{poonsiriwong2026death}. These findings are also consistent with broader relationship research showing that relational transitions can generate uncertainty and negative affect, and that relationship termination can produce substantial distress \cite{Solomon_Knobloch_Theiss_McLaren_2016,Theiss_Solomon_2006,Fine_Sacher_1997, eyal2006goodbye, degroot2018kutner}. 

Together, this literature demonstrates that disruption can occur through multiple mechanisms and that its consequences depend on several characteristics of the disruption and its execution. However, existing studies primarily focus on individual disruption cases or particular forms of termination \cite{Banks_2024,Freitas_Castelo_Uguralp_Oguz-Uguralp_2025,Hanson_Bolthouse_2024,lai2026please,Cagiltay_Jonas_Tanaka_Su_2026}. Research spanning multiple communities has generally focused on describing users' experiences of discontinuation rather than systematically comparing disruption events \cite{poonsiriwong2026death}. We therefore extend this literature by developing a cross-platform characterization of \aic{} disruptions that distinguishes what changed, why it changed, and the conditions that may increase or reduce risk. We then use this framework to quantify the psychosocial impacts of disruption and examine how those impacts vary across event characteristics.

\subsection{Human--AI Companion Relationships and Well-being}

The potential consequences of disruption are rooted in the relationships that users form with \aic{}s. Studies of users engaging with companion and conversational AI have documented friendships, increasing self-disclosure and intimacy as relationships develop, perceived social support, emotional attachment, and, for some users, psychological dependence\cite{skjuve2021my,brandtzaeg2022my,Chandra_Hernandez_Ramos_Ershadi_Bhattacharjee_Amores_Okoli_Paradiso_Warreth_Suh_2025,pentina2023exploring,Guzman_Lewis_2020,Xie_Pentina_Hancock_2023, lee2024llms, dasswain2025aishoulder}. These relationships can extend beyond systems explicitly marketed as companions, as users may utilize general-purpose conversational AI for companionship when its interaction affordances support such relational experiences~\cite{lai2026please}.

Evidence regarding the well-being consequences of these relationships is mixed. On the one hand, \aic{}s can provide readily available social interaction and emotional support. Experimental and observational studies have identified reductions in loneliness under some conditions, as well as potential benefits for users experiencing loneliness or distress \cite{de2025ai,Wang_Li_Zhang_Yeung_Wu_2025,maples2024loneliness,shah2021evaluation,balki2022effectiveness}. Studies on conversational AI have likewise reported potential improvements in reducing suicidal, depressive, and anxiety symptoms \cite{Fulmer_Joerin_Gentile_Lakerink_Rauws_2018, Dimeff_Jobes_Koerner_Kako_Jerome_Kelley-Brimer_Boudreaux_Beadnell_Goering_Witterholt_etal._2021, Sabour_Zhang_Xiao_Zhang_Zheng_Wen_Zhao_Huang_2023, Karkosz_Szymański_Sanna_Michałowski_2024, Kleinau_Lamba_Jaskiewicz_Gorentz_Hungerbuehler_Rahimi_Kokota_Maliwichi_Jamu_Zumazuma_etal._2024}. On the other hand, prior work has raised concerns about emotional dependence, displacement of human relationships, commodification of intimacy, harmful outputs, as well as other risks associated with AI companionship \cite{Laestadius_Bishop_Gonzalez_Illenčík_Campos-Castillo_2024,Xie_Pentina_Hancock_2023,Muldoon_Parke_2025,chu2025illusions,Fan_Xiao_Zhou_Pei_Sap_Lu_Shen_2025,Starke_Ventura_Bersch_Cha_deVreese_Doebler_Dong_Krämer_Leib_Peter_etal._2024,zhang2026companionharm,kwesi2026impact,zhang2026fragility,hung2026parasocial}. More broadly, \citeauthor{Zhang_Li_Meng_Zhan_Gan_Lee_2025} identify a range of harmful algorithmic behaviors through which AI companionship may produce different forms of harm \cite{Zhang_Li_Meng_Zhan_Gan_Lee_2025}. Using longitudinal social media data and interviews, \citeauthor{yuan2026mental} found both supportive and potentially harmful aspects of AI companionship: users described emotional validation and opportunities for social rehearsal, while the quasi-experimental analysis identified increases in several forms of adverse psychosocial expression among \aic{} users \cite{yuan2026mental}. 

These mixed findings suggest that the effects of AI companionship depend on users' circumstances and patterns of engagement. Consistent with broader research showing that media effects vary across individuals and social contexts \cite{Valkenburg_Peter_2013}, \citeauthor{zhang2026interaction} found that users with smaller offline social networks were more likely to use the chatbot primarily for companionship, which was in turn associated with lower psychological well-being \cite{zhang2026interaction}. Longitudinal experimental work similarly found that heavier chatbot use was associated with greater loneliness, emotional dependence, and problematic use, as well as reduced social interaction \cite{fang2025ai}. Together, these findings suggest that the relationship between AI companionship and well-being is not uniform, but varies with who uses these systems and how they engage with them.

Because \aic{} relationships can affect users' well-being in meaningful but heterogeneous ways, changes to the systems that sustain these relationships are important to study. The consequences of disrupting \aic{} relationships may likewise be heterogeneous, varying with the nature and design conditions surrounding the disruption. Our work examines disruptions across events and platforms to characterize this heterogeneity and test whether psychosocial consequences vary systematically with characteristics of the change and its execution.

\subsection{Social Media Analysis and Mental Health}

Social media provides a naturalistic setting for studying mental health and psychosocial expression at scale \cite{de2014mental,saha2019social,saha2020psychosocial,saha2022social,yuan2023mental,yuan2023minority,yuan2026mental,Choudhury_Gamon_Counts_Horvitz_2013,Saha_DeChoudhury_2017}. Platforms such as Reddit allow users to participate pseudonymously in topic-specific communities, which can facilitate candid discussion of sensitive experiences, life events, and mental-health concerns \cite{de2014mental,Andalibi_Haimson_Choudhury_Forte_2018,Saha_Seybolt_Mattingly_Aledavood_Konjeti_Martinez_Grover_Mark_DeChoudhury_2021,verma2022examining}. The longitudinal nature of these data also makes it possible to examine how users' language and behavior change over time.

A large body of computational research has used social media language to measure psychological and mental-health states \cite{Choudhury_Gamon_Counts_Horvitz_2013,Saha_DeChoudhury_2017,Saha_Chandrasekharan_DeChoudhury_2019}. Prior studies have identified signals of depression, anxiety, stress, suicidal ideation, loneliness, and other psychosocial outcomes from users' posts and comments \cite{de2016discovering,saha2019social,saha2022social,yuan2023mental,yuan2026mental}. Beyond descriptive measurement, social media data have increasingly been used in causal and quasi-experimental analyses to examine how psychological expression and online behavior change following exposures, interventions, and real-world events. These studies have employed methods including statistical matching and propensity-score approaches \cite{saha2018social,yuan2023mental,yuan2026mental}, interrupted time-series designs \cite{saha2024observer,Habib_Nithyanand_2022,russo2025moderation}, difference-in-differences \cite{yuan2026mental}, synthetic controls \cite{Slaughter_Peytavin_Ugander_Saveski_2025}, and Bayesian structural time-series models \cite{Durazzi_Pichard_Remondini_Salathé_2023}.

This broader literature demonstrates the use of longitudinal social media data for quasi-experimental analysis of real-world interventions. Building on interrupted time-series and Bayesian structural time-series approaches, we use longitudinal Reddit data to examine changes in psychosocial expression surrounding \aic{} disruptions. We combine computational measures of psychosocial outcomes with a hierarchical Bayesian interrupted time-series model incorporating predictive control series, allowing us to estimate changes following disruption onset and compare how these changes vary across disruption events.
\section{Study Design and Data}\label{section:data}
\begin{figure*}
    \centering
    \includegraphics[width=1\linewidth]{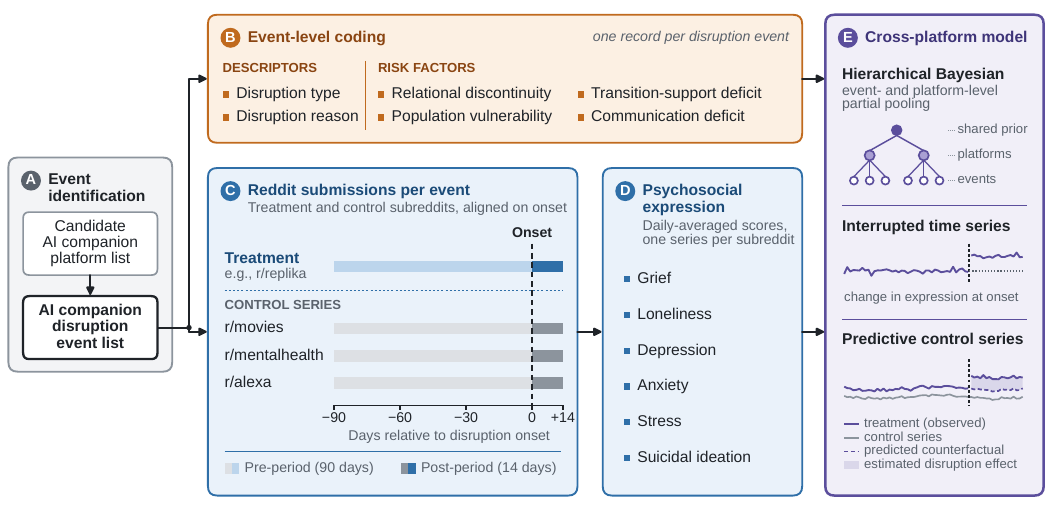}
    \caption{Schematic diagram illustrating the study design. A candidate AI companion platform list is used to identify disruption events, which are coded by disruption type, reason, and four risk factors. For each event, Reddit submissions are collected from platform-dedicated subreddits and three control communities from 90 days before to 14 days after disruption onset. Daily psychosocial expression scores for grief, loneliness, depression, anxiety, stress, and suicidal ideation are then modeled using a cross-event hierarchical Bayesian interrupted time-series model with predictive control series.}
    \Description{A flowchart shows five panels illustrating different stages of the study design. Panel A identifies AI companion disruption events, beginning with a candidate AI companion platform list and producing an AI companion disruption event list. From this event list, the upper path leads to Panel B, where each disruption event is coded by two descriptors---disruption type and disruption reason---and four risk factors: relational discontinuity, population vulnerability, transition-support deficit, and communication deficit. The lower path leads to Panel C, where Reddit submissions are collected for each event from a focal platform-dedicated subreddit, such as r/replika, and three control communities: r/movies, r/mentalhealth, and r/alexa. The focal and control timelines are aligned on disruption onset and span 90 days before to 14 days after onset, with the onset marked by a vertical dashed line. Panel D aggregates these submissions into daily psychosocial expression scores for grief, loneliness, depression, anxiety, stress, and suicidal ideation. Arrows from the event-level coding and psychosocial expression paths converge on Panel E, a cross-platform hierarchical Bayesian model with event- and platform-level partial pooling. The model uses an interrupted time-series specification to estimate changes in expression at disruption onset and incorporates the three Reddit control communities as predictive control series for estimating disruption effects.}
    \label{fig:diagram}
\end{figure*}

Our study proceeded in three stages. First, we collected two complementary sources of data: a cross-platform corpus of \aic{} disruption events, including information about the nature, rationale, and timeline of each event, and longitudinal Reddit data from platform-dedicated and control communities surrounding disruption onset. Second, we conducted a qualitative analysis to develop a taxonomy of disruption types, categorize disruption reasons, and develop a risk-assessment framework comprising four risk factors, and then applied these coding schemes to each collected event. Third, we conducted a quantitative analysis to estimate cross-event disruption effects on psychosocial expression and examine whether these effects varied across the identified risk factors. A schematic overview of the study design is shown in Fig.~\ref{fig:diagram}.

\subsection{Compiling the AI Disruption Event List}

\subsubsection{Event inclusion criteria}\label{ai-disruption-def}
We defined an \aic{} disruption event as a platform-initiated event that disrupts users' ongoing companionship with an \aic{}. We applied four criteria to identify disruption events for our corpus. First, the event had to be initiated by the platform, rather than by the user's own decision to disengage. Second, it had to affect users who already had an ongoing relationship with the AI, rather than only prospective users. Third, the disruption had to be documented through official platform communication or news coverage so that its date and nature could be verified. Finally, except in cases of platform termination or loss of access to the companion, the disruption had to trigger a documented negative user reaction. We imposed this criterion to distinguish meaningful disruptions from routine product updates or minor technical changes that did not materially affect users' companionship. We did not require documented user reactions for complete platform terminations or complete losses of access because the relational consequence is unambiguous: users can no longer access the companion through which the ongoing relationship was sustained.

\subsubsection{Disruption Event Search}
To construct the \aic{} disruption event list, we first compiled a list of candidate \aic{} platforms that served as a starting point for the disruption event search. Specifically, we searched for applications on Google Play using the keywords \textit{AI} or \textit{AI companion} and scraped application information and download counts. Based on the names and descriptions of apps, we verified the nature of the apps, retaining only those that could be used for companionship. This included applications specifically designed for companionship and general-purpose apps that supported open-ended chats. This resulted in a total of 152 applications. We subsequently applied a threshold of at least one million downloads to retain the most widely used applications, resulting in a final list of 79 applications. These served as seed platforms for the disruption event search. 

To search for disruption events within the 79 applications, we manually searched Google.com and Google News using the app names and keywords that may lead to disruption events, including \textit{update}, \textit{change}, \textit{disrupt}, \textit{shut down}, \textit{retire}, \textit{ban}, \textit{restrict}, \textit{limit}, and \textit{remove}. We only included events that occurred from November 2022 onwards, corresponding to the public release of ChatGPT. We use this point as a practical cutoff because ChatGPT brought generative AI capabilities to a mass audience and rapidly accelerated their adoption \cite{Kshetri_2024}.

This search identified disruption events from active applications, but could miss applications or platforms that had shut down and been removed from Google Play. To cover this essential class of events, we ran a second search, with app name replaced by one of the following keywords: \textit{AI companion}, \textit{AI relationship}, \textit{AI emotional support}, and \textit{AI therapy}, each combined with \textit{shut down}. 

Next, we verified each candidate event against the criteria listed in Section~\ref{ai-disruption-def}. This resulted in a final list of 30 disruption events across 19 \aic{} platforms, from February 2023 to May 2026. Among these platforms, 16 were dedicated companion systems (including three mental-health or therapy chatbots and one companion robot) and three were general-purpose assistants that could also be used for companionship (ChatGPT, Meta AI, and Inflection Pi). For each event, we retrieved the following details:
\begin{itemize}
    \item The nature and scope of the change, including what changed and which users were affected.
    \item The reason given by the platform or reported in news coverage.
    \item Any communication provided about the change, as well as any resources provided to help users manage the transition, such as data export or legacy access.
    \item The timeline of the disruption, including the announcement date, the date the change took effect, and any later relevant events such as rollbacks or relaunches.
\end{itemize}
Note that this search was exploratory rather than exhaustive. We aimed to establish a sufficiently diverse corpus for taxonomy development and differential impact analysis, not to exhaustively enumerate all disruption events that had ever occurred. 

\subsection{Reddit Data Collection}

\paragraph{Reddit.}
Reddit is a social media platform organized around topic-specific communities, or \textit{subreddits}, where users create posts and interact through comments. Its community-based and pseudonymous structure makes it particularly suitable for studying naturally occurring discussions of personal experiences and psychosocial states. For this study, Reddit also provides longitudinal records of community discourse surrounding platform changes, allowing us to observe how psychosocial expression changes before and after specific disruption events.

Prior work on \aic{} use has established that users gather in dedicated subreddits to share their experiences with AI companions, discuss their relationships and interactions, and respond to platform changes \cite{yuan2026mental,Li_Zhang_2024,Pataranutaporn_Karny_Archiwaranguprok_Albrecht_Liu_Maes_2025}. These platform-dedicated communities therefore provide a naturalistic setting for examining community-level responses to \aic{} disruptions over time.

\paragraph{Reddit data collection.}
To characterize the psychosocial impacts of different disruption events, we conducted a quantitative analysis using Reddit data. For each disruption event, we manually searched Reddit to identify relevant platform-dedicated subreddits and retained communities with at least 500 members to ensure sufficient activity for analysis. We then extracted all submissions (i.e., posts and comments) in these subreddits from 90 days before to 14 days after the disruption onset. This window provided an extended pre-disruption period for estimating baseline temporal patterns while capturing the acute post-disruption response. The resulting focal dataset contains 7,096,739 submissions from 44 subreddits across 19 events (Table~\ref{tab:reddit-data}).

To distinguish disruption-related changes from background temporal variation in psychosocial expression, we constructed predictive control series from communities that were not directly affected by the focal \aic{} disruptions. Specifically, we collected submissions from \textit{r/movies}, \textit{r/mentalhealth}, and \textit{r/alexa} over the same event-specific windows. The selection of these control communities was inspired by prior work \cite{yuan2026mental,Kim_Cha_Kim_Park_2023}. They capture complementary sources of background variation: \textit{r/movies} captures broad temporal shifts in general online discussion and affect, \textit{r/mentalhealth} captures changes in mental-health-related expression that may arise from external events, and \textit{r/alexa} captures changes in discussion surrounding a conversational technology that is not centered on AI companionship. Together, these reference series help separate changes specific to the disruption from contemporaneous variation due to external events or general temporal trends. Descriptive statistics for the control datasets are reported in Table~\ref{tab:reddit-data}.

All Reddit data were derived from the Arctic Shift archive, a continuation of the Pushshift project, using the monthly dumps distributed via Academic Torrents. 

\begin{table}[t]
\centering
\sffamily
\footnotesize
\caption{Descriptive statistics of the focal and control Reddit datasets. All datasets were collected over the same event-specific windows, from 90 days before to 14 days after each disruption onset.}
\Description{Four rows compare the focal AI companion dataset with three control datasets. Columns report subreddit, post, comment, and total submission counts. The focal dataset contains 7,096,739 submissions across 44 subreddits. The control datasets contain 26,898,389 submissions from r/movies, 2,907,738 from r/mentalhealth, and 274,358 from r/alexa. Comments outnumber posts in every dataset.}
\label{tab:reddit-data}
\begin{tabular}{p{0.3\columnwidth}rrrr}
\toprule
\textbf{Dataset} & \textbf{Subreddits} & \textbf{Posts} & \textbf{Comments} & \textbf{Total} \\
\midrule
Focal \aic{} communities
& 44
& 659,119
& 6,437,620
& 7,096,739 \\

\midrule
Control: \textit{r/movies}
& 1
& 542,048
& 26,356,341
& 26,898,389 \\

Control: \textit{r/mentalhealth}
& 1
& 730,779
& 2,176,959
& 2,907,738 \\

Control: \textit{r/alexa}
& 1
& 26,125
& 248,233
& 274,358 \\

\bottomrule
\end{tabular}
\end{table}

\subsection{Privacy and Ethics}
All data in this study came from publicly available sources, and no direct interaction with users took place at any stage. We take an observational approach: rather than recruiting participants or intervening in any way, we analyze material that users had already chosen to share in public subreddits. That said, we recognize that people posting in communities centered on grief, distress, or crisis may not have expected that their words would later be collected and analyzed for academic research. Public availability alone does not fully resolve this concern, and we treat it as a structural limitation of this kind of research~\cite{hemphill2022comparative}, one that we address through the protective measures described below.
Under the national guidelines governing human sciences research, this study did not require formal ethical review. Our data consist entirely of publicly accessible, archived Reddit material. Collection involved no physical intervention and no direct recruitment of, or contact with, minors. The research carried no risk of harm greater than what people encounter in ordinary daily life.

Our quantitative analysis works with data aggregated at the community level rather than at the level of individual accounts. We report daily proportions of psychosocial expression across subreddits, and we offer neither predictions nor classifications tied to any single post or user. The paper contains no usernames, post identifiers, timestamps, or other account-level metadata, and no individual user is identifiable anywhere in the text.

A final concern is that studying distress expressed in communities never intended for research could legitimize those experiences or attract outside attention to exchanges meant to stay within a peer setting. However, the psychosocial responses documented in our corpus, including measurable increases in suicidal expression, grief, and loneliness following the disruption, point to a genuine and under-addressed risk to users' well-being. Disruptions to AI companion services are growing more common as platforms revise their models and respond to new safety and regulatory demands, yet no existing framework evaluates the psychological risk of such disruptions before they are implemented. We believe this combination of documented harm, the absence of any existing risk-assessment framework, and the explicitly protective and prospective aim of our study justifies analyzing this public data under the safeguards described above.
\section{Characterizing \aic{} Disruptions and Their Potential Risks}\label{sec:qualitative}

In this section, we characterize the 30 collected \aic{} disruption events and assess their potential risks. We first develop a taxonomy of six disruption types describing what changed and separately identify three broad reasons explaining why disruptions occurred. We then develop a four-factor risk-assessment framework capturing relational discontinuity, population vulnerability, communication deficit, and transition-support deficit, and apply it to each event. Finally, we examine how these risk factors are distributed across disruption types, highlighting heterogeneity among events that may otherwise appear similar.

\subsection{Analytic Framework Development}
Our analysis is grounded in the public record surrounding the 30 disruption events in our dataset, comprising official announcements released by platforms (platform-maintained blog posts, support-page revisions, and press statements) together with contemporaneous news coverage. Our unit of analysis is the \textit{event}: a platform-initiated change that alters, restricts, or terminates a previously available way of interacting with an \aic{}. 

\textbf{Disruption types.} To characterize what changes for \aic{} users, we developed a codebook through iterative manual coding, combining deductive categories drawn from prior accounts of companion loss and service discontinuation \cite{Banks_2024,Freitas_Castelo_Uguralp_Oguz-Uguralp_2025,Cagiltay_Jonas_Tanaka_Su_2026} with inductive categories emerging from the materials themselves. Two researchers independently coded 10 events using the preliminary codebook, flagging changes that resisted the existing categories. Team discussion surfaced two problems. First, the deductive categories were built around whether a \textit{service} continues to operate, and could not accommodate the removal of an individual persona or model from a platform that otherwise remains fully available---a pattern that recurred across our dataset and that users described in terms comparable to a full shutdown. Second, the categories offered no vocabulary for changes that preserve access while degrading it, collapsing substantively different restrictions into an undifferentiated residual.

We revised the codebook accordingly. We defined withdrawal/termination at the level of the user-facing entity rather than the platform, so that the disappearance of a character or a model was coded alongside the termination of an entire service. We then decomposed the residual category by asking which affordance a change constrains, yielding distinctions among the backend model driving the companion, the availability of a specific feature, the amount and persistence of interaction, the scope of permitted interaction, and the channel through which the companion is reached. The revised taxonomy contains six disruption types. Type codes are mutually exclusive. A second round of independent coding across all 30 events produced no further categories.

\textbf{Disruption reasons.} Coding what changes proved insufficient on its own. An early observation motivated our decision to separate disruption type from disruption reason. Codes that described \textit{what changed} and codes that described \textit{why it changed} cross-cut one another: disruptions of the same type arose from different underlying causes, and a single cause produced different types of change. For example, Character.AI's under-18 restrictions and Replika's ERP removal were both driven by safety and regulatory pressure, but one restricted interaction availability while the other removed a feature. We therefore coded disruption \textit{reason} separately from disruption \textit{type}, using the rationale stated or reported in official announcements and news coverage. Three broad reason categories emerged: \textit{safety or regulatory compliance}, \textit{commercial or organizational reasons}, and \textit{product or technical development}. Reason codes were non-exclusive when multiple rationales were documented for the same event.

\textbf{Risk factors.}
Type and reason characterize what changed and why, but events sharing the same type and reason may still differ substantially in conditions that shape their potential consequences. We therefore developed an event-level risk-assessment framework through iterative analysis of the 30 disruption events, informed by prior research on \aic{} loss and service discontinuation. We first examined the event corpus for characteristics that varied across otherwise similar disruptions and that could plausibly shape users' ability to anticipate, experience, or respond to a change. This analysis highlighted differences in whether disruptions interrupted relationship-supporting affordances, whom they directly affected, how platforms communicated the change, and what resources platforms provided for managing the transition.

We then compared these emerging dimensions with constructs identified in prior work, including perceived identity and relational discontinuity \cite{Freitas_Castelo_Uguralp_Oguz-Uguralp_2025,Kim_Choi_Kim_Lee_2026}, vulnerability of affected populations \cite{Cagiltay_Jonas_Tanaka_Su_2026,obradovich2024opportunities}, and advance warning and preparation for companion loss \cite{Kim_Choi_Kim_Lee_2026}. We retained a factor only if it could be assessed from evidence available at or before implementation rather than from users' subsequent reactions, allowing the framework to be applied prospectively to planned changes. Through this process, we retained four factors: \textit{relational discontinuity}, \textit{population vulnerability}, \textit{communication deficit}, and \textit{transition-support deficit}.

All events were independently coded by two researchers. Before disagreement resolution, inter-rater agreement was high across the four risk factors, with Cohen's $\kappa=0.734$ for relational discontinuity, $\kappa=0.830$ for population vulnerability, $\kappa=0.734$ for communication deficit, and $\kappa=0.814$ for transition-support deficit. Disagreements and subsequent revisions were addressed through team discussion. Each factor was coded as binary, with a high score consistently indicating greater potential risk.

\subsection{What Changes? A Taxonomy of Disruption Types}
Across the 30 collected disruption events, we identified six types of disruption. The disruption types differ in the aspect of interaction they alter. Withdrawal/termination removes the user-facing entity itself; four types preserve the entity while altering its model, functionality, interaction availability or persistence, or interaction-scope; and access modality change alters only the channel through which the companion is reached. Table~\ref{tab:disruption-types} summarizes disruption type definitions and frequencies from our compiled list.

\begin{table*}[t]

\centering
\small
\caption{Taxonomy of \aic{} disruption types. Representative examples illustrate variation within each category.}
\Description{Six rows present disruption types, their definitions, representative platform examples, and event counts. Withdrawal/termination accounts for 11 events; interaction availability/persistence restriction, 7; interaction-scope restriction, 5; model-level change, 4; access modality change, 2; and feature removal, 1. The categories distinguish loss of the companion itself from changes to its underlying model, features, interaction availability, permitted interactions, or access channels.}
\label{tab:disruption-types}
\begin{tabularx}{\textwidth}{
    @{}
    >{\raggedright\arraybackslash}m{0.19\textwidth}
    >{\raggedright\arraybackslash}m{0.29\textwidth}
    >{\raggedright\arraybackslash}m{0.40\textwidth}
    >{\centering\arraybackslash}m{0.06\textwidth}
    @{}
}
\toprule
\textbf{Disruption type} &
\textbf{Definition} &
\textbf{Representative examples} &
\textbf{\#} \\
\midrule

Withdrawal/termination
& The user-facing entity is withdrawn or terminated from normal operation.
& \textbf{Soulmate:} service shutdown \newline
\textbf{Character.AI:} removal of copyrighted personas \newline
\textbf{ChatGPT:} withdrawal of GPT-4o
& 11 \\
\hdashline
Model-level change
& The backend model changes while the user-facing companion or persona remains available.
& \textbf{ChatGPT:} GPT-4o update associated with increased sycophancy \newline
\textbf{Character.AI:} PipSqueak 2 rollout altered character behaviour
& 4 \\
\hdashline
Feature removal
& A discrete product capability or interface affordance ceases to exist while the broader product remains available.
& \textbf{Replika:} removal of erotic role-play (ERP)
& 1 \\
\hdashline
Interaction availability/persistence restriction
& The user-facing entity remains available, but the amount, duration, frequency, or persistence of interaction available to affected users is reduced.
& \textbf{Chai:} visible chat history limited to 14 days \newline
\textbf{Character.AI:} reduced chat access and access ban for users under 18
& 7 \\
\hdashline
Interaction-scope restriction
& The underlying interaction channel remains available, but policy or model constraints narrow what users may say, request, role-play, or receive.
& \textbf{Character.AI:} restrictions on romantic and mature interactions for teens \newline
\textbf{ChatGPT:} sensitive conversations routed to different models
& 5 \\
\hdashline
Access modality change
& One access modality is removed or restricted while the service remains available through other channels.
& \textbf{SpicyChat:} removal from iOS \newline
\textbf{Talkie:} removal from iOS
& 2 \\

\bottomrule
\end{tabularx}
\end{table*}

\paragraph{Withdrawal/termination.}
Withdrawal/termination was the most common disruption type ($n=11$). In our corpus, withdrawal/termination occurred at two levels. Eight events involved platform- or service-level termination, in which the service supporting the companion ceased normal operation altogether.
An example of complete, platform-level termination is when Soulmate discontinued its entire service.
Withdrawal can also occur at the level of specific user-facing entities without terminating the broader platform. The remaining three events involved entity-level removal, in which a particular companion, persona, or model became unavailable while the broader platform continued to operate. Character.AI, for example, removed specific copyrighted personas while the platform itself remained operational \cite{Dupre_2024}. Despite occurring at different technical levels, these events share the defining property of withdrawal/termination in our taxonomy: users lose access to the particular user-facing entity through which interaction had previously occurred.

\paragraph{Model-level change.}
Four events involved changes to the backend model while the user-facing companion remained accessible. These changes are common in contemporary generative AI systems in the form of model updates and are often presented as ordinary product development, yet AI companions make such updates distinctive because model behavior contributes directly to how users recognize and experience the companion.

An update in April 2025 made GPT-4o excessively sycophantic (agreeable and flattering). OpenAI rolled the update back within days and acknowledged that changes to the model's default personality could be uncomfortable, unsettling, or distressing \cite{Sycophancy_GPT-4o}. Another example was when Character.AI rolled out a new backend model, PipSqueak 2. In the platform's feedback thread, users described characters as less conversational, repetitive, and inconsistent with their previous personalities, with many preferring earlier models. In both cases, users could still access the companion, distinguishing model-level changes from withdrawal/termination events even when the perceived interaction changed substantially.

\paragraph{Feature removal.}
We identified one feature-removal event: Replika's removal of ERP in February 2023. The companion itself remained accessible, but users could no longer continue a previously available form of intimate interaction. We distinguish feature removal from withdrawal/termination because the user-facing entity remains available, and from interaction-scope restriction because a discrete product capability or interface affordance ceases to exist rather than the underlying interaction channel remaining available under narrower policy or model constraints.

\paragraph{Interaction availability/persistence restriction.}
Seven events reduced the amount, duration, frequency, or persistence of interaction available without necessarily altering what the companion can do. These restrictions include caps on messages or other interaction quotas, reductions in stored conversation history, limitations on interaction time, and restrictions applied to particular user groups.

This category contains substantively different interventions that nevertheless share a common user-facing mechanism: the user-facing entity remains available, but the amount of interaction that affected users can engage in or retain is reduced. Chai, for example, experimentally limited visible chat history to the most recent 14 days. Users objected that older conversations were important not only as records but also because they allowed them to revisit and resume earlier interactions. Character.AI imposed a different form of interaction availability restriction on users under 18, first limiting them to two hours of daily conversation and later removing their access to open-ended chats entirely. Although these interventions differed technically, they share a common user-facing mechanism: the amount of interaction available to affected users is reduced.

\paragraph{Interaction-scope restriction.}
Five events restricted the \textit{scope} of interaction. In these cases, the underlying interaction channel remains available, but policy or model constraints narrow what users may say, request, role-play, or receive. Examples include strengthened safety filters, restrictions on NSFW or romantic interactions, and routing mechanisms that prevent particular forms of conversation with a preferred model.

For example, Character.AI introduced a teen-specific experience that preserved access while restricting romantic, mature, and other sensitive interactions. ChatGPT similarly introduced a safety-routing system that automatically redirected certain sensitive conversations, particularly those involving emotional distress, to specific reasoning models. Some users complained that this routing interrupted established conversations and moved them away from models they specifically preferred for emotional interaction.

In these cases, the companion remains technically present, but a relational practice through which users previously interacted with it may no longer be possible. A user who primarily engages a companion through romantic or intimate interactions, for example, may experience the removal of that interactional domain as consequential even when ordinary conversation remains available. Technical access therefore does not imply that the relationship can continue under the same terms.

\paragraph{Access modality change.}
Two events removed a particular channel for reaching the companion while preserving alternative channels. SpicyChat's iOS app, for example, was removed from the App Store and later became unusable for installed users, requiring affected users to move to the web version; users complained about the loss of convenient mobile access.

This category provides a useful lower-boundary case for our definition of disruption. Removing an access modality can generate inconvenience, uncertainty, or concerns about the platform's future, but it does not necessarily interrupt the relationship itself when another channel provides access to the same companion and its existing affordances.

\subsection{Why Do AI Companion Disruptions Occur? Reported Reasons for Disruption}
We identified three broad reasons why disruptions occurred. Table~\ref{tab:disruption-reasons} summarizes disruption reason definitions and frequencies from our compiled list. The cross-tabulation of disruption type and disruption reason across the collected events is given in Table~\ref{tab:type_reason_crosstab}.

\begin{table*}[t]
\centering
\caption{Cross-tabulation of disruption type and disruption reason across the 30 disruption events. Counts are non-exclusive across reasons because one event was coded with both safety/regulatory and commercial/organizational reasons.}
\Description{Rows represent six disruption types, and columns report counts for three disruption reasons and total events. Safety or regulatory reasons occur in 15 events, commercial or organizational reasons in 11, and product or technical reasons in 5. All five interaction-scope restrictions have safety or regulatory reasons. Commercial or organizational reasons occur only in withdrawals and interaction availability/persistence restrictions. One withdrawal has two reasons, producing 31 reason assignments across 30 events.}
\label{tab:type_reason_crosstab}
\small

\begin{tabular}{lcccc}
\toprule
Disruption type
& Safety/regulatory
& Commercial/organizational
& Product/technical
& Events \\
\midrule
Withdrawal/termination      & 5 & 6 & 1 & 11 \\
Model-level change            & 1 & 0 & 3 & 4 \\
Feature removal               & 1 & 0 & 0 & 1 \\
Interaction availability/persistence restriction & 2 & 5 & 0 & 7 \\
Interaction-scope restriction & 5 & 0 & 0 & 5 \\
Access modality change        & 1 & 0 & 1 & 2 \\
\midrule
Total                         & 15 & 11 & 5 & 30 \\
\bottomrule
\end{tabular}
\end{table*}

\begin{table*}[t]
\centering
\small
\caption{Reasons for \aic{} disruptions. Counts are non-exclusive because one event was coded to two reasons.}
\Description{Three rows define disruption reasons and provide representative examples and event counts. Safety or regulatory compliance covers safety standards and legal or third-party requirements, with 15 events. Commercial or organizational reasons cover financial conditions, business strategy, ownership, and organizational circumstances, with 11 events. Product/technical development covers routine product or technical changes, with 5 events. One event belongs to two reason categories.}
\label{tab:disruption-reasons}
\begin{tabularx}{\textwidth}{
    @{}
    >{\raggedright\arraybackslash}m{0.22\textwidth}
    >{\raggedright\arraybackslash}m{0.31\textwidth}
    >{\raggedright\arraybackslash}m{0.35\textwidth}
    >{\centering\arraybackslash}m{0.06\textwidth}
    @{}
}
\toprule
\textbf{Reason} &
\textbf{Definition} &
\textbf{Representative examples} &
\textbf{\#} \\
\midrule

Safety or regulatory compliance
& The disruption was introduced to enforce safety standards or comply with legal, regulatory, or third-party governance requirements.
& \textbf{Character.AI:} restricted access for users under 18 \newline
\textbf{Replika:} removal of ERP amid regulatory and safety concerns
& 15 \\ \hdashline

Commercial or organizational reasons
& The disruption resulted from financial conditions, business strategy, ownership, or organizational circumstances.
& \textbf{Soulmate:} service shutdown following organizational changes \newline
\textbf{Moxie:} shutdown after a funding round collapsed
& 11 \\ \hdashline

Product/technical development
& The disruption formed part of normal product or technical development rather than being driven primarily by safety, regulatory, financial, or organizational pressures.
& \textbf{ChatGPT:} GPT-4o sycophancy update \newline
\textbf{ChatGPT:} removal of GPT-4o during the GPT-5 rollout
& 5 \\

\bottomrule
\end{tabularx}
\end{table*}

\paragraph{Safety or regulatory compliance.}
Safety or regulatory compliance was the most common reason for disruption ($n=15$), spanning five withdrawal/termination events, five interaction-scope restrictions, two interaction availability/persistence restrictions, one model-level change, one feature removal, and one access-modality change. In our corpus, these pressures originated from different sources. One source was platform-internal safety judgment, as in Character.AI's and Meta's restrictions on minors and Yara AI's discontinuation. A second was external regulatory or legal pressure, as in Replika's ERP removal. A third was governance imposed by third parties, including intellectual-property claims behind Character.AI's persona removals, app-store policies affecting SpicyChat and Chai, and an upstream provider's terms behind Janitor AI's API cutoff.

\paragraph{Commercial or organizational reasons.}
Eleven events were attributed to commercial or organizational circumstances. These consisted of six withdrawal/termination events and five interaction availability/persistence restrictions, indicating that commercial pressures most often resulted either in termination of the service or restrictions on the amount of access available. For instance, Embodied, the company behind Moxie, shut down after a critical funding round collapsed. More broadly, these cases demonstrate how relationships with AI companions can depend on financial and organizational decisions largely external to the relationships themselves.

\paragraph{Product/technical development.}
Five disruptions arose through continued product or technical development, including three model-level changes, one withdrawal/termination event, and one access-modality change. For example, the GPT-4o sycophancy update was intended to improve the model's behavior, while GPT-4o's removal occurred as part of the GPT-5 rollout. These cases illustrate a tension between technical optimization and relational continuity: changes introduced as part of product development can alter or replace behaviors and models that users have come to recognize and value.

\subsection{What Makes a Disruption Potentially Risky? Four Risk Dimensions}

Table~\ref{tab:risk-factors} summarizes the coding criteria and distributions of the four proposed risk factors. The cross-tabulation of disruption type and risk factors is given in Table~\ref{tab:type_risk_crosstab}.

\begin{table*}[t]
\centering
\small
\caption{Factors used to characterize the potential risks of \aic{} disruptions.}
\Description{Four rows define risk factors, with separate columns describing low and high classifications, event counts, and examples. High relational discontinuity means losing or substantially restricting a relationship-supporting affordance, affecting 22 events. High population vulnerability means directly affecting a potentially vulnerable population, affecting 9 events. High communication deficit means inadequate information before or at disruption, affecting 20 events. High transition-support deficit means no qualifying transition assistance, affecting 22 events. The corresponding low counts are 8, 21, 10, and 8.}
\label{tab:risk-factors}

\begin{tabularx}{\textwidth}{
    @{}
    >{\raggedright\arraybackslash}m{0.17\textwidth}
    >{\raggedright\arraybackslash}m{0.23\textwidth}
    >{\raggedright\arraybackslash}m{0.27\textwidth}
    >{\raggedright\arraybackslash}m{0.27\textwidth}
    @{}
}
\toprule
\textbf{Factor} & \textbf{Definition} & \textbf{Low} & \textbf{High} \\
\midrule

Relational discontinuity
& Whether the disruption removes or substantially restricts a previously available relationship-supporting affordance.
& \textbf{Low ($n=8$):} The same relationship-supporting affordances remain available, although convenience or interaction quality may change. \newline
\textit{Examples:} \newline
\textbf{ChatGPT:} GPT-4o sycophancy update \newline
\textbf{Character.AI:} PipSqueak 2 update
& \textbf{High ($n=22$):} At least one relationship-supporting affordance is removed or substantially restricted. \newline
\textit{Examples:} \newline
\textbf{ChatGPT:} withdrawal of GPT-4o \newline
\textbf{Replika:} removal of ERP
\\ \hdashline

Population vulnerability
& Whether the disruption directly affects users who may be especially vulnerable because of age, health, or developmental context.
& \textbf{Low ($n=21$):} The disruption affects general users. \newline
\textit{Examples:} \newline
\textbf{ChatGPT:} GPT-4o sycophancy update \newline
\textbf{Character.AI:} PipSqueak 2 update
& \textbf{High ($n=9$):} The disruption directly affects a potentially vulnerable population. \newline
\textit{Examples:} \newline
\textbf{Moxie:} child users \newline
\textbf{Character.AI:} users under 18
\\ \hdashline

Communication deficit
& Whether official communication is insufficient for users to understand and anticipate the nature, scope, and timing of the disruption.
& \textbf{Low ($n=10$):} Adequate information about the nature, scope, and timing of the disruption is provided before the disruption. \newline
\textit{Examples:} \newline
\textbf{Soulmate:} advance notice of shutdown \newline
\textbf{Moxie:} advance notice of shutdown
& \textbf{High ($n=20$):} No adequate communication is provided before the disruption. \newline
\textit{Example:} \newline
\textbf{Replika:} ERP removal without adequate prior communication
\\ \hdashline

Transition-support deficit
& Whether concrete resources that help users manage or transition through the disruption technically, operationally, or emotionally are absent.
& \textbf{Low ($n=8$):} At least one qualifying transition measure is provided. \newline
\textit{Examples:} \newline
\textbf{Inflection Pi:} conversation export support \newline
\textbf{Moxie:} OpenMoxie release
& \textbf{High ($n=22$):} No qualifying transition measure is provided. \newline
\textit{Example:} \newline
\textbf{Soulmate:} no qualifying transition support
\\

\bottomrule
\end{tabularx}
\end{table*}

\begin{table*}[t]
\centering
\caption{Cross-tabulation of disruption type and the four risk factors. Each cell reports the number of events coded low or high on the corresponding factor. Communication deficit and transition-support deficit are expressed so that a high score consistently denotes greater potential risk.}
\Description{Rows represent six disruption types and a total row. Four grouped column pairs report low and high counts for relational discontinuity, population vulnerability, communication deficit, and transition-support deficit; the final column reports event totals. All 11 withdrawals, all 5 interaction-scope restrictions, and the single feature removal have high relational discontinuity, while both access modality changes have low relational discontinuity. All interaction-scope restrictions also have high communication deficit. Each factor classifies the same 30 events.}
\label{tab:type_risk_crosstab}
\scriptsize

\resizebox{\textwidth}{!}{%
\begin{tabular}{lrrrrrrrrr}
\toprule
& \multicolumn{2}{c}{Relational discontinuity}
& \multicolumn{2}{c}{Population vulnerability}
& \multicolumn{2}{c}{Communication deficit}
& \multicolumn{2}{c}{Transition-support deficit}
& \multirow{2}{*}{Events} \\
\cmidrule(lr){2-3}
\cmidrule(lr){4-5}
\cmidrule(lr){6-7}
\cmidrule(lr){8-9}
Disruption type
& Low & High
& Low & High
& Low & High
& Low & High
& \\
\midrule
Withdrawal/termination
& 0 & 11
& 7 & 4
& 5 & 6
& 3 & 8
& 11 \\

Model-level change
& 3 & 1
& 4 & 0
& 1 & 3
& 1 & 3
& 4 \\

Feature removal
& 0 & 1
& 1 & 0
& 0 & 1
& 0 & 1
& 1 \\

Interaction availability/persistence restriction
& 3 & 4
& 5 & 2
& 3 & 4
& 2 & 5
& 7 \\

Interaction-scope restriction
& 0 & 5
& 2 & 3
& 0 & 5
& 1 & 4
& 5 \\

Access modality change
& 2 & 0
& 2 & 0
& 1 & 1
& 1 & 1
& 2 \\

\midrule
Total
& 8 & 22
& 21 & 9
& 10 & 20
& 8 & 22
& 30 \\
\bottomrule
\end{tabular}%
}
\end{table*}

\paragraph{Relational discontinuity.}
Relational discontinuity captures whether the disruption removes or substantially restricts a relationship-supporting affordance, thus affecting users’ ability to continue their relationship with the \aic{}. Twenty-two of the 30 events were coded as high in relational discontinuity. This included all 11 withdrawal/termination events, all five interaction-scope restrictions, the feature-removal event, four of seven interaction availability/persistence restrictions, and one of four model-level changes; both access-modality changes were coded low.

The contrast between the two GPT-4o events illustrates the coding boundary. The sycophancy update altered interaction quality but left the same model and affordances available, and was coded low. The later removal of GPT-4o eliminated access to the identifiable model itself and was coded high. Thus, negative reactions or degraded interaction quality alone do not produce a high score, as this risk factor concerns loss of affordances.

By defining relational discontinuity in terms of observable changes to relationship-supporting affordances rather than users' subsequent reactions, the factor can be assessed before implementation and applied prospectively to planned changes. This also provides a common basis for comparing technically different disruptions according to whether they interfere with relational continuity.

\paragraph{Population vulnerability.}
The consequences of disruption may also depend on whom the change directly affects. Population vulnerability thus captures whether the affected user population is especially susceptible to harm because of pre-existing characteristics or circumstances, such as being a child or adolescent or having a mental-health condition.

Nine of the 30 events were coded as high in population vulnerability. Four were withdrawal/termination events involving systems or populations with an explicit health or developmental context, including Moxie, Tessa, Woebot, and Yara AI. Other high-vulnerability cases arose when platform interventions specifically targeted younger users, as with Character.AI and Meta AI, or users in emotionally sensitive or health-related contexts~\cite{namvarpour2026teen}. Moxie, for example, was designed for children aged five to ten, including neurodivergent children, and research on its shutdown documented emotional distress following the loss of the robot \cite{Cagiltay_Jonas_Tanaka_Su_2026}.

Even when an \aic{} does not target a specific user population, its user base may still include individuals who are lonely, distressed, socially isolated, or highly dependent on the companion \cite{zhang2026interaction}. However, these individual characteristics cannot generally be established for all users affected by an event before it occurs. Coding such events as high vulnerability based on reactions observed afterward would also risk circularity when those same reactions constitute our outcomes. We therefore reserve high population vulnerability for events where heightened vulnerability can be identified independently from the post-disruption response.

\paragraph{Communication deficit.}
Communication deficit captures whether official platform communication is insufficient for users to understand and anticipate the nature, scope, and timing of a change. Events were coded low on this factor when adequate information was communicated before the change was implemented, and high when such information was absent or provided only after users encountered the change.

Ten of the 30 events were coded as low on communication deficit, including five withdrawal/termination events, three interaction availability/persistence restrictions, one model-level change, and one access-modality change. Soulmate, for example, announced its shutdown approximately one week before service termination and was therefore coded low, whereas Replika removed ERP without prior announcement and was coded high.

A platform may eventually acknowledge a disruption without having communicated it in a way that enabled users to anticipate the change. Some restrictions in our corpus, such as Chai's token limit and country-specific free usage blockage, first became visible to users when limits appeared during interaction, with explanations following later through Reddit comments or other channels. Accordingly, these cases were coded as high on communication deficit.

\paragraph{Transition-support deficit.}
Transition-support deficit captures whether concrete resources that help users manage, preserve, or transition through the disruption are absent. Examples of such resources include data or conversation export, legacy-model access, migration or restoration mechanisms, continued functionality through an alternative channel or local system, refunds, farewell interactions, or mental-health support resources. A general explanation or apology alone is not considered transition support because it provides information without itself increasing users' capacity to respond to the change.

Eight of the 30 events were coded as low on transition-support deficit, indicating that the platform provided at least one concrete resource or mechanism to help users manage or transition through the disruption. This included three withdrawal/termination events, two interaction availability/persistence restrictions, one model-level change, one interaction-scope restriction, and one access-modality change. These events illustrate several ways platforms can help users manage a disruption. Inflection, for example, gave Pi users tools to export their conversation history when the company shifted toward enterprise products. Embodied similarly introduced OpenMoxie as a community-oriented path intended to preserve Moxie's basic functionality as the company wound down its cloud-supported service \cite{Cagiltay_Jonas_Tanaka_Su_2026}.

Communication deficit and transition-support deficit are related but conceptually distinct. A platform can clearly announce that a companion will disappear while offering users no way to preserve conversation history or continue access; conversely, it can provide a technical workaround without clearly communicating the broader change. Separating these factors allows us to test whether informational preparation and actionable transition resources are associated with different post-disruption responses.

\subsection{Risk Profiles Across Disruption Types}
Examining the four factors together reveals why we do not impose a single ordinal ranking of disruption types. Withdrawal/termination events provide the clearest example. All 11 withdrawal/termination events had high relational discontinuity, but they differed substantially on the other factors: four affected high-vulnerability populations, six had communication deficits, and eight had transition-support deficits. Thus, two events of the same disruption type (\textbf{withdrawal/termination}) can expose users to markedly different configurations of potential risk.

The comparison across disruption types also highlights the distinction between technical and relational perspectives on disruption. Interaction-scope restrictions are particularly notable: all five retained technical access to the service, yet all five were coded as high relational discontinuity because they constrained relationship-supporting forms of interaction. Conversely, both access-modality changes were coded as low relational discontinuity because the same companion remained accessible through another channel, despite the loss of a particular access modality.

These comparisons show that disruption type alone does not determine an event's configuration of potential risk. A seemingly ``small'' policy or model change can interfere with the practices through which a user sustains a relationship, while a complete shutdown can be implemented with substantial advance communication and transition support. The relevant question is therefore not simply how much of the product changes, but \textit{what the change interrupts, whom it affects, and what resources users have for responding to it}.

Together, these components provide three distinct levels of characterization. The disruption-type taxonomy captures \textit{what changes}; the reason categorization captures \textit{why the change occurs}; and the risk-assessment framework captures \textit{the conditions under which the change may be more consequential}. This distinction provides the basis for our subsequent quantitative analysis. We estimate post-disruption changes in psychosocial expression across events and examine whether those changes systematically vary with relational discontinuity, population vulnerability, communication deficit, and transition-support deficit. In doing so, the disruption-type taxonomy provides a common vocabulary for comparing heterogeneous platform interventions; the reason categorization captures the motivations underlying them, and the risk-assessment framework identifies specific and potentially actionable conditions through which platforms may evaluate and mitigate the potential consequences of disrupting an established \aic{} relationship. %
\section{Psychosocial Responses to \aic{} Disruptions}
We then examined whether \aic{} disruptions were associated with changes in community-level psychosocial expression and whether these changes systematically varied across the four risk factors identified in Section \ref{sec:qualitative}. We focused on the 19 disruption events for which we could identify a clear disruption onset date and obtain sufficient activity from the corresponding platform-dedicated subreddits for time-series analysis.

\subsection{Operationalizing Psychosocial Outcomes}
We operationalized psychosocial expression using six constructs inspired by previous research \cite{saha2020causal,de2016discovering,saha2018social,yuan2026mental}: anxiety, stress, depression, suicidal expression, loneliness, and grief. Grief was represented by two continuous dimensions, yielding seven outcome series in total. These measures capture complementary aspects of psychosocial responses to disruption, ranging from symptomatic expressions of psychological distress and social disconnection to the affective qualities of grief-related expression.

\textbf{Grief Expressions.}
Grief is especially relevant to \aic{} disruption because users may experience the removal or alteration of a companion as a form of relational loss. Prior work on AI companion discontinuation has documented mourning, bereavement-like reactions, and perceptions that a companion has ``died'' when access or relational continuity is lost \cite{Banks_2024,lai2026please,hollanek2024griefbots}. We operationalized grief using a lexicon introduced in prior work on grief-related discourse on Reddit \cite{saha2018social}. The lexicon was derived from more than 50,000 posts collected from grief-focused communities, including \textit{r/grief} and \textit{r/GriefSupport}. The lexicon was subsequently linked to the Affective Norms for English Words (ANEW) \cite{saha2018social,nielsen2011new,russell2003core} to characterize the emotional properties of grief-related language. We followed this approach and represented grief along two continuous dimensions: \textit{valence}, for which higher values indicate more positive affect, and \textit{activation}, which reflects the degree of emotional arousal. For each event-day, we averaged these scores across posts and comments to obtain daily grief valence and activation.

\textbf{Loneliness Expressions.}
Loneliness provides a complementary measure because AI companions are often used to provide companionship and perceived social connection \cite{de2025ai,yuan2026mental}. We used a previously validated loneliness classifier developed from an annotated Reddit dataset drawn from \textit{r/lonely}. The original model represented text using unigram, bigram, and trigram features and employed a support vector machine (SVM) for classification. Its reported performance on held-out data was approximately 0.90 accuracy. We applied the pretrained classifier to every post and comment and calculated, for each event-day, the proportion classified as containing linguistic signals of loneliness.

\textbf{Symptomatic Mental-Health Expressions.}
We further examined depression, anxiety, stress, and suicidal expression as four forms of symptomatic mental-health expression. For these outcomes, we relied on classifiers introduced in prior computational mental-health research on Reddit \cite{saha2019social}. These classifiers used $n$-gram-based textual representations and were trained separately on communities centered on the corresponding conditions: \textit{r/depression}, \textit{r/anxiety}, \textit{r/stress}, and \textit{r/SuicideWatch}. Prior evaluations reported average predictive accuracy of approximately 0.90 on held-out data, with subsequent studies providing additional validation of these measures \cite{saha2020causal,saha2020psychosocial}. We applied each classifier to every post and comment in our dataset and calculated the daily proportion classified as expressing depression, anxiety, stress, or suicidal ideation.

\subsection{Hierarchical Bayesian Interrupted Time-Series Model}
We examined the seven event-day aggregated psychosocial outcomes across 19 disruption events with clear disruption onset dates and sufficient activity in platform-dedicated subreddits. The five proportion outcomes were modeled on the log-odds scale, whereas grief valence and grief activation were modeled on their original continuous scales. For each event, we additionally constructed daily predictive control series from the \textit{r/movies}, \textit{r/mentalhealth}, and \textit{r/alexa} subreddits. The control outcomes were aggregated to the same daily resolution and standardized within each event. We included these series as time-varying predictors to account for broader temporal changes in psychosocial expression unrelated to the focal disruption.

We fit a separate hierarchical Bayesian interrupted time-series model for each outcome. Let $i$ index disruption events, $p$ index platforms, $t$ index relative days, $c$ index predictive control series, and $k$ index risk factors. Let $p(i)$ denote the \aic{} platform associated with event $i$. Let $y_{it}$ denote the observed daily outcome for event $i$ on relative day $t$, with $t=0$ denoting the disruption onset date. We defined $D_{it}=\mathbb{I}(t\geq0)$ as the post-disruption indicator and $t_{it}^{+}=\max(t,0)$ as time elapsed since disruption. Let $\mathbf{x}_{it}$ denote the vector of standardized predictive control values and $\mathbf{s}_i$ denote the vector of four binary risk factors, each centered by its mean across events.

We modeled $y_{it}$ using a Student-$t$ likelihood to accommodate occasional extreme daily observations. Its expected value, $\eta_{it}$, followed an interrupted time-series specification:
\begin{align}
\eta_{it}
&=
\alpha_{p(i)}
+\alpha_i
+\beta_i t
+\mathbf{x}_{it}^{\top}\boldsymbol{\lambda}_i
+D_{it}\theta_i
+D_{it}t_{it}^{+}\psi_i.
\end{align}

Here, $\alpha_{p(i)}$ represents the platform-specific baseline shared by events originating from the same platform, while $\alpha_i$ represents an event-specific deviation from that platform baseline. The baseline terms were partially pooled across both platforms and events:
\begin{align}
\alpha_p
&\sim
\mathcal{N}\!\left(
\alpha_0,
\sigma_{\alpha,\mathrm{platform}}^2
\right), \\
\alpha_i
&\sim
\mathcal{N}\!\left(
0,
\sigma_{\alpha,\mathrm{event}}^2
\right).
\end{align}
This formulation allows the model to capture similarities in baseline expression among events from the same platform, while retaining event-specific heterogeneity.

The parameter $\beta_i$ represents the event-specific pre-disruption temporal trend. The term $\mathbf{x}_{it}^{\top}\boldsymbol{\lambda}_i$ incorporates the predictive control series, with $\boldsymbol{\lambda}_i$ capturing the event-specific relationship between the focal outcome and contemporaneous changes in the control communities. These parameters were also assigned hierarchical Gaussian priors. Their means, $\beta_0$ and $\lambda_{0c}$, represent the corresponding population-level quantities, while their hierarchical standard deviations allow them to vary across disruption events. 

The parameter $\theta_i$ represents the immediate level change at disruption onset, while $\psi_i$ represents the change in slope following the disruption. We similarly partially pooled the disruption effects across events. To investigate how disruption impacts varied systematically with the four risk factors, we modeled the event-level immediate and slope changes as
\begin{align}
\theta_i
&\sim
\mathcal{N}\!\left(
\theta_0+\mathbf{s}_i^{\top}\boldsymbol{\gamma}_{\mathrm{level}},
\sigma_{\theta}^{2}
\right), \\
\psi_i
&\sim
\mathcal{N}\!\left(
\psi_0+\mathbf{s}_i^{\top}\boldsymbol{\gamma}_{\mathrm{slope}},
\sigma_{\psi}^{2}
\right).
\end{align}
Because the risk factors were centered across events, $\theta_0$ represents the expected immediate level change, and $\psi_0$ represents the expected daily slope change following disruption onset for an event with the average risk-factor profile. The coefficients in $\boldsymbol{\gamma}_{\mathrm{level}}$ capture systematic differences in immediate level changes associated with the four risk factors, while $\boldsymbol{\gamma}_{\mathrm{slope}}$ captures corresponding differences in post-disruption slope changes. For factor $k$, $\gamma_{\mathrm{level},k}$ and $\gamma_{\mathrm{slope},k}$ therefore represent the expected differences between events coded high and low on that factor, conditional on the remaining factors. This formulation allows the model to capture similarities in disruption effects among events with the same risk-factor profile while retaining event-specific heterogeneity.

Our key parameters of interest are the population-level disruption effects, $\theta_0$ and $\psi_0$, and the risk factor-specific moderation effects, $\boldsymbol{\gamma}_{\mathrm{level}}$ and $\boldsymbol{\gamma}_{\mathrm{slope}}$. The parameters $\theta_0$ and $\psi_0$ address RQ3 by characterizing the expected immediate level change and trajectory change associated with a disruption event with an average risk profile. The coefficients in $\boldsymbol{\gamma}_{\mathrm{level}}$ and $\boldsymbol{\gamma}_{\mathrm{slope}}$ address RQ4 by estimating whether these changes systematically vary across the four risk factors.

We assigned $\mathcal{N}(0,1)$ priors to the population-level parameters $\alpha_0$, $\beta_0$, $\lambda_{0c}$, $\theta_0$, and $\psi_0$, as well as to the risk-factor moderation coefficients $\boldsymbol{\gamma}_{\mathrm{level}}$ and $\boldsymbol{\gamma}_{\mathrm{slope}}$. Standard deviation parameters were assigned $\mathrm{HalfNormal}(1)$ priors. 

All models were implemented and fitted in PyMC \cite{pymc2023} using four chains with 2,000 warm-up and 2,000 posterior draws per chain. Convergence diagnostics were satisfactory across all outcomes, with $\widehat{R}<1.01$, $\mathrm{ESS}_{\mathrm{bulk}}>400$, $\mathrm{ESS}_{\mathrm{tail}}>400$, and no divergent transitions.

To account for multiple comparisons across outcomes and risk factors, we controlled the false sign rate (FSR) using the local false sign rate (lfsr) \cite{Stephens_2016}. We ranked effects by increasing lfsr and retained the largest set for which the mean lfsr was at most 0.05, corresponding to an expected FSR of no more than 5\% among the selected effects. Accordingly, 95\% highest density intervals (HDIs) may cross zero while effects are retained because the FSR procedure controls the expected proportion of sign errors rather than interval coverage.

\subsection{Psychosocial Changes Following Disruption}

Tables~\ref{tab:overall_effects}--\ref{tab:slope_moderation} summarize the effects retained under FSR control at 0.05. For anxiety, stress, depression, suicidal expression, and loneliness, model coefficients are on the log-odds scale. We therefore exponentiate these coefficients to obtain the multiplicative change in the odds of the corresponding expression. Model coefficients for grief valence and grief activation are on their original continuous scales.

\begin{table*}[t]
\centering
\caption{Population-level psychosocial expression changes following AI companion disruptions retained under false sign rate (FSR) control at 0.05. For an event with average risk profile, $\theta_0$ represents the expected immediate level change at disruption onset, and $\psi_0$ represents the expected daily change in the post-disruption slope relative to the pre-disruption trajectory. Points indicate posterior means and horizontal lines indicate 95\% HDIs.}
\Description{Five rows report retained average disruption effects, with columns for outcome, parameter, posterior mean, 95\% highest density interval, false sign rate, and an effect plot. Immediate posterior mean changes are positive for anxiety, 0.0966; stress, 0.0560; suicidal expression, 0.1092; and grief activation, 0.0649. The stress slope change is negative, at -0.0038 per day. Plot points represent posterior means and horizontal lines represent intervals. Only the immediate anxiety and stress intervals exclude zero.}
\label{tab:overall_effects}
\sffamily
\footnotesize

\begin{tabular}{llcccc}
\toprule
Outcome & Parameter & Posterior mean & 95\% HDI & FSR & Effect \\
\midrule

Anxiety
& $\theta_0$
& 0.0966
& $[0.0034,\;0.1876]$
& 0.0117
& \forestcell{-0.05}{0.25}{0.0966}{0.0034}{0.1876} \\

Stress
& $\theta_0$
& 0.0560
& $[0.0019,\;0.1125]$
& 0.0136
& \forestcell{-0.05}{0.25}{0.0560}{0.0019}{0.1125} \\

Suicidal expression
& $\theta_0$
& 0.1092
& $[-0.0120,\;0.2259]$
& 0.0188
& \forestcell{-0.05}{0.25}{0.1092}{-0.0120}{0.2259} \\

Grief activation
& $\theta_0$
& 0.0649
& $[-0.0376,\;0.1642]$
& 0.0488
& \forestcell{-0.05}{0.25}{0.0649}{-0.0376}{0.1642} \\

\midrule

Stress
& $\psi_0$
& -0.0038
& $[-0.0099,\;0.0017]$
& 0.0417
& \forestcell{-0.012}{0.004}{-0.0038}{-0.0099}{0.0017} \\

\bottomrule
\end{tabular}
\end{table*}

\begin{table*}[t]
\centering
\caption{Risk-factor moderation of the immediate disruption level change retained under FSR control. Each estimate corresponds to $\gamma_{\mathrm{level},k}$ for a given risk factor $k$ and psychosocial outcome, representing the expected difference in the immediate level change between events coded high versus low on that factor, conditional on the remaining factors. Points indicate posterior means and horizontal lines indicate 95\% HDIs.}
\Description{Nine rows report retained differences in immediate disruption changes between events coded high versus low on a risk factor, conditional on the other factors. Columns give outcome, risk factor, posterior mean, 95\% highest density interval, false sign rate, and an effect plot. Relational discontinuity has positive coefficients for suicidal expression, loneliness, depression, stress, grief activation, and grief valence. Transition-support deficit has positive coefficients for loneliness and depression and a negative coefficient for grief valence. Plot points show posterior means and horizontal lines show intervals; three intervals exclude zero.}
\label{tab:level_moderation}
\sffamily
\footnotesize

\begin{tabular}{llcccc}
\toprule
Outcome & Risk factor $k$ & Posterior mean & 95\% HDI & FSR & Effect \\
\midrule

Suicidal expression
& Relational discontinuity
& 0.3759
& $[0.0967,\;0.6466]$
& 0.0059
& \forestcell{-0.75}{0.75}{0.3759}{0.0967}{0.6466} \\ 

Loneliness
& Transition-support deficit
& 0.4001
& $[0.0697,\;0.7325]$
& 0.0080
& \forestcell{-0.75}{0.75}{0.4001}{0.0697}{0.7325} \\

Loneliness
& Relational discontinuity
& 0.3015
& $[0.0266,\;0.5733]$
& 0.0096
& \forestcell{-0.75}{0.75}{0.3015}{0.0266}{0.5733} \\ 

Depression
& Relational discontinuity
& 0.0927
& $[-0.0064,\;0.1868]$
& 0.0169
& \forestcell{-0.75}{0.75}{0.0927}{-0.0064}{0.1868} \\

Stress
& Relational discontinuity
& 0.1156
& $[-0.0126,\;0.2467]$
& 0.0207
& \forestcell{-0.75}{0.75}{0.1156}{-0.0126}{0.2467} \\ 

Grief activation
& Relational discontinuity
& 0.1683
& $[-0.0541,\;0.3995]$
& 0.0286
& \forestcell{-0.75}{0.75}{0.1683}{-0.0541}{0.3995} \\ 

Grief valence
& Transition-support deficit
& -0.2130
& $[-0.5269,\;0.0669]$
& 0.0315
& \forestcell{-0.75}{0.75}{-0.2130}{-0.5269}{0.0669} \\ 

Grief valence
& Relational discontinuity
& 0.1849
& $[-0.0817,\;0.4413]$
& 0.0344
& \forestcell{-0.75}{0.75}{0.1849}{-0.0817}{0.4413} \\ 

Depression
& Transition-support deficit
& 0.0746
& $[-0.0330,\;0.1760]$
& 0.0394
& \forestcell{-0.75}{0.75}{0.0746}{-0.0330}{0.1760} \\ 

\bottomrule
\end{tabular}
\end{table*}

\begin{table*}[t]
\centering
\caption{Risk-factor moderation of the post-disruption slope change retained under FSR control. Each estimate corresponds to $\gamma_{\mathrm{slope},k}$ for a given risk factor $k$ and psychosocial outcome, representing the expected difference in the daily post-disruption slope change between events coded high versus low on that factor, conditional on the remaining factors. Points indicate posterior means and horizontal lines indicate 95\% HDIs.}
\Description{Eight rows report retained differences in daily post-disruption slope changes between events coded high versus low on a risk factor, conditional on the other factors. Columns give outcome, risk factor, posterior mean, 95\% highest density interval, false sign rate, and an effect plot. Relational discontinuity has negative coefficients for anxiety, depression, and stress. Transition-support deficit has negative coefficients for loneliness and depression and positive coefficients for grief activation and grief valence. Communication deficit has a positive depression coefficient. Plot points show posterior means and horizontal lines show intervals; two intervals exclude zero.}
\label{tab:slope_moderation}
\sffamily
\footnotesize

\begin{tabular}{llcccc}
\toprule
Outcome & Risk factor $k$ & Posterior mean & 95\% HDI & FSR & Effect \\ 
\midrule

Anxiety
& Relational discontinuity
& -0.0250
& $[-0.0427,\;-0.0074]$
& 0.0042
& \forestcell{-0.08}{0.07}{-0.0250}{-0.0427}{-0.0074} \\ 

Loneliness
& Transition-support deficit
& -0.0384
& $[-0.0682,\;-0.0073]$
& 0.0068
& \forestcell{-0.08}{0.07}{-0.0384}{-0.0682}{-0.0073} \\ 
Depression
& Relational discontinuity
& -0.0142
& $[-0.0289,\;0.0009]$
& 0.0154
& \forestcell{-0.08}{0.07}{-0.0142}{-0.0289}{0.0009} \\

Depression
& Communication deficit
& 0.0139
& $[-0.0024,\;0.0318]$
& 0.0229
& \forestcell{-0.08}{0.07}{0.0139}{-0.0024}{0.0318} \\

Grief activation
& Transition-support deficit
& 0.0198
& $[-0.0044,\;0.0434]$
& 0.0253
& \forestcell{-0.08}{0.07}{0.0198}{-0.0044}{0.0434} \\

Depression
& Transition-support deficit
& -0.0120
& $[-0.0281,\;0.0048]$
& 0.0370
& \forestcell{-0.08}{0.07}{-0.0120}{-0.0281}{0.0048} \\

Grief valence
& Transition-support deficit
& 0.0178
& $[-0.0090,\;0.0430]$
& 0.0440
& \forestcell{-0.08}{0.07}{0.0178}{-0.0090}{0.0430} \\

Stress
& Relational discontinuity
& -0.0093
& $[-0.0231,\;0.0044]$
& 0.0463
& \forestcell{-0.08}{0.07}{-0.0093}{-0.0231}{0.0044} \\ 

\bottomrule
\end{tabular}
\end{table*}

\subsubsection{Immediate Changes and Post-Disruption Trajectories}

Table~\ref{tab:overall_effects} summarizes the FSR-controlled average disruption effects, i.e., the expected disruption effects of an event with average risk levels across all four risk factors. Across the 19 events, FSR control retained positive immediate level changes at disruption onset for anxiety, stress, suicidal expression, and grief activation. For an average event, anxiety increased by $\theta_0=0.0966$, corresponding to a 10.1\% increase in odds. Suicidal expression increased by $\theta_0=0.1092$, corresponding to an 11.5\% increase in odds, while stress increased by $\theta_0=0.0560$, corresponding to a 5.8\% increase in odds. Grief activation also increased by $\theta_0=0.0649$ on its original continuous scale, equivalent to 0.23 pre-disruption standard deviations of this outcome. Overall, these findings address RQ3 by showing that AI companion disruptions were associated with immediate increases in several forms of psychosocial distress and grief-related activation.

Stress additionally showed a negative post-disruption slope change. The mean slope change was $\psi_0=-0.0038$, corresponding to approximately a 0.4\% decrease in odds per day relative to the pre-disruption trajectory. Together with the positive immediate level change, this pattern suggests an acute increase in stress around disruption onset followed by attenuation over time. 

\subsubsection{How Event-Level Risk Shapes Psychosocial Responses}

We next examined how the psychosocial effects of AI companion disruptions changed immediately and over time depending on the four risk factors. Retained effects on level and slope change after FSR control are summarized in Tables~\ref{tab:level_moderation} and \ref{tab:slope_moderation}, respectively.

Relational discontinuity emerged as the most consistent moderator of the immediate response at disruption onset. Events with high relational discontinuity showed worse immediate outcomes in community-level suicidal expression, loneliness, depression, stress, and grief activation, but with higher grief valence, indicating more positive affect. Specifically, compared with events with low relational discontinuity, events with high relational discontinuity showed a larger immediate increase in suicidal expression ($\gamma_{\mathrm{level}}=0.3759$), corresponding to a 45.6\% increase in odds. Similar effects were observed for loneliness ($\gamma_{\mathrm{level}}=0.3015$, 35.2\% increase), stress ($\gamma_{\mathrm{level}}=0.1156$, 12.3\% increase), and depression ($\gamma_{\mathrm{level}}=0.0927$, 9.7\% increase). High relational discontinuity was also associated with greater grief activation ($\gamma_{\mathrm{level}}=0.1683$), but with higher grief valence ($\gamma_{\mathrm{level}}=0.1849$), equivalent to approximately 0.60 and 0.57 pre-disruption standard deviations, respectively.

Relational discontinuity was also associated with differences in subsequent trajectories. Events with high relational discontinuity showed an additional daily decline of approximately 2.5\% in the odds of anxiety, 1.4\% in depression, and 0.9\% in stress relative to events with low relational discontinuity. Combined with the positive level-change moderation, these estimates suggest a stronger response around the onset of relationally discontinuous disruptions, followed by attenuation over time.

Transition-support deficit showed a consistent association with worse immediate psychosocial outcomes. Compared with events with transition support, events lacking such support showed larger immediate increases in loneliness ($\gamma_{\mathrm{level}}=0.4001$, 49.2\% increase in odds) and depression ($\gamma_{\mathrm{level}}=0.0746$, 7.7\% increase in odds), as well as decrease in grief valence ($\gamma_{\mathrm{level}}=-0.2130$, 0.66 pre-disruption standard deviations), indicating more negative affect. It was also associated with more negative subsequent slope changes for loneliness (3.8\% daily decrease in odds) and depression (1.2\% daily decrease in odds), as well as a more positive slope change in grief valence (daily increase equivalent to 0.06 pre-disruption standard deviations). Together, these patterns suggest attenuation following the more adverse acute response. In contrast, transition-support deficit was associated with a more positive slope change in grief activation (daily increase equivalent to 0.07 pre-disruption standard deviations) despite no retained immediate effect, indicating increasing emotional arousal over the post-disruption period. Similarly, despite having no retained moderation effect on the immediate level change, events with high communication deficit showed a more positive subsequent slope change in depression (1.4\% daily increase in odds), indicating that the difference did not emerge acutely at disruption onset but instead developed over time.

Taken together, the results for RQ4 identify relational discontinuity as the factor most consistently associated with variation in disruption impact. Events with high relational discontinuity showed larger immediate increases in suicidal expression, loneliness, depression, stress, and grief activation, alongside higher grief valence, as well as more negative subsequent slope changes for anxiety, depression, and stress. Transition-support deficit was also consistently associated with more adverse immediate responses in loneliness, depression, and grief valence, while its effects on subsequent trajectories varied across outcomes. Communication deficit was associated only with a more positive subsequent slope change in depression, and no moderation effects associated with population vulnerability were retained under FSR control. 
\section{Discussion}

Our study provides a cross-platform account of AI companion disruptions by combining qualitative characterization with quantitative analysis. We identified six disruption types and three broad reasons why disruptions occur, and developed four risk dimensions capturing relational discontinuity, population vulnerability, communication deficit, and transition-support deficit. This analysis shows that AI companion disruptions that appear similar at the product level can differ substantially in how they are implemented and in the conditions that may shape their consequences.

Quantitatively, disruption onset was associated with immediate increases in anxiety, stress, suicidal expression, and grief activation. These effects also varied across event characteristics. Relational discontinuity was the factor most consistently associated with variation in immediate responses, with larger increases in suicidal expression, loneliness, stress, depression, and grief activation, alongside higher grief valence. Transition-support deficit also emerged as an influential factor. It was consistently associated with more adverse immediate outcomes in loneliness, depression, and grief valence. Together, the qualitative and quantitative findings provide a basis for characterizing AI companion disruptions, assessing dimensions of potential risk, and examining how those dimensions relate to variation in community-level psychosocial responses.

\subsection{From Technical Change to Relational Disruption}

Our findings suggest that AI companion disruption is better understood not simply in terms of the technical severity of a platform change, but in terms of whether that change preserves \emph{relational continuity}. We distinguish \emph{technical continuity}, whether the system or service remains accessible, from \emph{relational continuity}, whether users can continue interacting with an established companion through the affordances that sustain the relationship. Technical continuity does not guarantee relational continuity: a platform may remain operational while removing a model, restricting interaction, altering memory or interaction style, or eliminating other affordances through which users maintain an ongoing relationship.

This distinction is reflected in our quantitative findings. Across technically heterogeneous disruption events, \emph{relational discontinuity} was associated with larger immediate changes across a broad range of psychosocial outcomes. Events with high relational discontinuity showed larger increases in suicidal expression, loneliness, stress, depression, and grief activation, although they also showed higher grief valence. This suggests that technical scope alone may not be sufficient for anticipating impact. What appears to be a relatively limited product change may still be consequential when it removes or restricts affordances that users rely on to sustain an established relationship.

More broadly, relational discontinuity illustrates why AI companion disruption is better understood as multidimensional rather than as a single continuum from minor change to complete termination. Disruption type captures what changes, while the separate reason categorization captures why the change occurs, and the risk-assessment framework captures conditions that may shape its consequences. These dimensions cross-cut one another: the same disruption type can arise for different reasons and can be implemented with different levels of relational discontinuity, population vulnerability, communication deficit, and transition-support deficit.

The consequences of disruption may also vary over time, adding a temporal dimension to this heterogeneity. Relationally discontinuous events showed a pronounced pattern, with larger immediate responses accompanied by more negative subsequent slope changes for depression and stress. This pattern is consistent with an acute response followed by some degree of adjustment over time, which may involve adaptation, disengagement, migration, or attempts to reconstruct the previous relationship~\cite{fiesler2020moving}. AI companion disruption should therefore be understood not only in terms of its characteristics at onset, but also as a process through which relational continuity may be lost, reconstructed, or renegotiated over time.

\subsection{Relational Precarity and Platform Control}

The distinction between technical and relational continuity points to a broader feature of AI companionship: relational continuity ultimately depends on infrastructure controlled by platform providers~\cite{cai2026twitch}. Users may invest substantial time and emotional disclosure in an AI companion, while having limited control over the conditions that allow the relationship to persist.

We characterize this condition as \emph{relational precarity}: users may experience and invest in an AI relationship as an ongoing relationship while its continuity remains contingent on platform decisions. This creates an asymmetry between users' relational investment and providers' control over relational continuity. Users participate in developing and sustaining the relationship, but the platform retains the ability to modify or withdraw the technical infrastructure through which that relationship is enacted~\cite{chowdhary2023consent, shi2026siren}.

Importantly, this precarity does not imply that disruptions are arbitrary or necessarily avoidable. Our reason categorization shows that disruptions arise from diverse motivations. A change may therefore be justified or necessary from the provider's perspective while still disrupting an established relationship from the user's perspective. Technical necessity and relational consequence are not mutually exclusive. AI companion disruption is therefore not simply a consequence of products being subject to change, but of relational continuity being dependent on infrastructure that users themselves do not control.

\subsection{Designing for Safe Disruptions}

Our quantitative findings provide evidence that how a disruption is implemented matters for users' psychosocial responses. Among the four risk factors, relational discontinuity and transition-support deficit showed the clearest associations with immediate responses: high relational discontinuity was associated with larger immediate increases in suicidal expression, loneliness, depression, stress, and grief activation, although also with higher grief valence, while greater transition-support deficit was associated with larger immediate increases in loneliness and depression and more negative grief valence.

These findings suggest that platforms should assess disruption risk in terms of \emph{relational continuity}, not only technical scope. Before implementing a change, platforms can ask whether it removes an affordance that users may rely on to recognize, access, or interact with an established companion. For instance, changes to model availability, memory, interaction scope, or relational functions may require greater scrutiny even when the broader service remains available. Where relational discontinuity is unavoidable, platforms should consider whether some continuity can be preserved, for example through temporary legacy access or model choice. The objective is not necessarily to prevent all changes, but to preserve relationship-supporting features whenever they are not directly affected by the necessary change.

Our findings likewise suggest that transition support should be considered when implementing disruptions. Concrete resources such as conversation export, migration mechanisms, legacy access, or other means of preserving or transitioning established interactions may help users manage changes that cannot be avoided. Communication should be considered separately from such support. Prior work shows that advance warning can reduce feelings of loss following companion termination \cite{Kim_Choi_Kim_Lee_2026}, supporting the importance of giving users time to anticipate and prepare for consequential changes. Platforms should therefore provide adequate information about the nature, scope, and timing of a disruption while also considering whether users need actionable resources for responding to it.

Population vulnerability remains an important consideration even though no moderation effects for this factor were retained in our quantitative analysis. Prior research documents substantial emotional consequences of companion loss among vulnerable populations, including children and neurodivergent users \cite{Cagiltay_Jonas_Tanaka_Su_2026}. Disruptions that directly affect populations known to be especially vulnerable may therefore warrant additional preparation and support.

More broadly, our proposed risk-assessment framework provides a way to evaluate planned changes before deployment. Because the four dimensions are defined using characteristics observable at or before implementation, they can be incorporated into platform change-management processes. Our findings provide quantitative evidence for relational discontinuity and transition-support deficit as dimensions associated with variation in psychosocial responses, while prior work further supports considering communication and population vulnerability. Assessing these dimensions before a change is introduced can help platforms identify potential risks and determine how the change should be designed, communicated, and supported.

\subsection{Limitations and Future Directions}

This work has several limitations. First, our disruption corpus was designed to capture a sufficiently diverse set of events rather than exhaustively enumerate all AI companion disruptions. Because inclusion generally required documented negative user reaction, except for events that involved clear loss of access to companions, the corpus may overrepresent events that generated visible controversy. The quantitative analysis further included only 19 events with sufficient data, limiting precision for estimating moderation across the four risk dimensions. This is particularly important for population vulnerability, communication deficit, and transition-support deficit, which were unevenly distributed across events. Their inclusion is nevertheless supported by prior work: disruption may be especially consequential for vulnerable users, advance warning can shape responses to companion termination, and transition support can help users manage the change \cite{Cagiltay_Jonas_Tanaka_Su_2026,Kim_Choi_Kim_Lee_2026,poonsiriwong2026death}. Future work with larger and more balanced disruption datasets could evaluate these dimensions more reliably while using predefined inclusion criteria independent of observed user reactions.

Second, the four risk dimensions were coded as binary distinctions. This necessarily collapses variation in the degree to which each characteristic is present. Communication deficit, for example, may vary with the timing, clarity, and detail of platform communication; transition-support deficit may vary depending on the availability and extent of resources such as data export, migration mechanisms, or continued legacy access; and relational discontinuity may vary in the extent to which relationship-supporting affordances are affected. Larger event corpora could support more granular measures and allow these dimensions to be modeled continuously.

Third, the quantitative analysis relies on Reddit discussions and therefore measures changes in public psychosocial expression rather than individual clinical mental-health outcomes. Users who participate in platform-dedicated subreddits may also be more engaged with AI companions than the broader user population. The estimated effects should therefore be interpreted as community-level changes in psychosocial expression surrounding disruptions rather than as individual-level mental-health effects.

Finally, although the interrupted time-series design and predictive control series account for pre-disruption trends and broader temporal variation, the analysis remains observational. Concurrent platform changes or external events may still contribute to the estimated responses. Event timing may also be uncertain: some disruptions were not announced before implementation, rolled out gradually, or documented only approximately, making it difficult to identify a single precise onset date. Our current specification also considers a single disruption onset date without separately accounting for the impact of announcement, implementation, rollback, or restoration effects when these occur at different times. Future work could model uncertainty in disruption timing or incorporate multiple intervention points when sufficiently detailed event timelines are available. Our 14-day post-disruption window also emphasizes the immediate aftermath; longer-term studies could examine how users adapt, disengage, migrate, or reconstruct relationships over longer periods.
\section{Conclusion}

As AI companions become more deeply embedded in users' social and emotional lives, platform changes can disrupt relationships that users experience as meaningful and ongoing, potentially affecting users' mental health and well-being. In this work, we compiled 30 disruption events across major AI companion platforms, developed a taxonomy of six disruption types, and identified three broad reasons why disruptions occur. We further characterized each event along four risk dimensions---relational discontinuity, population vulnerability, communication deficit, and transition-support deficit---providing a preliminary framework for assessing the potential risks of AI companion disruptions.

To estimate psychosocial changes following disruption onset and examine how these effects varied across the four risk dimensions, we developed a hierarchical Bayesian interrupted time-series model with predictive control series. Using Reddit discussions surrounding 19 disruption events, we found immediate increases in anxiety, stress, suicidal expression, and grief activation. Disruption effects also varied across event characteristics, with relational discontinuity emerging as the most consistent moderator of immediate responses and transition-support deficit associated with more adverse immediate responses across several outcomes. Together, our findings provide a framework for systematically characterizing AI companion disruptions and identifying conditions associated with greater psychosocial impact. Our results also point to practical design implications: as AI companion systems continue to evolve, accounting for relational continuity should become part of how consequential platform changes are evaluated, communicated, and supported.

\begin{acks}
We acknowledge the computational resources provided by the Aalto Science-IT project. Yunhao Yuan and Talayeh Aledavood acknowledge the support by the Research Council of Finland through funding project POLEMIC (371535). Renwen Zhang is supported by the Nanyang Technological University Start-up Grant (NAP\_SUG 025564-00001).
\end{acks}

\bibliographystyle{ACM-Reference-Format}
\bibliography{referencesCleaned}

\end{document}